\documentclass[a4paper,
               ]{jacow}
\usepackage{pdfpages,multirow,ragged2e, subfig} %
\makeatletter%
	\ifboolexpr{bool{xetex}}
	 {\renewcommand{\Gin@extensions}{.pdf,%
	                    .png,.jpg,.bmp,.pict,.tif,.psd,.mac,.sga,.tga,.gif,%
	                    .eps,.ps,%
	                    }}{}
\makeatother

\ifboolexpr{bool{xetex} or bool{luatex}} 
 {}                                      
 {\usepackage[utf8]{inputenc}}           

\usepackage[USenglish]{babel}

\ifboolexpr{bool{jacowbiblatex}}%
 {%
  \addbibresource{jacow-test.bib}
  \addbibresource{biblatex-examples.bib}
 }{}
\begin{document}

\title{Transverse Phase Space Tomography at FACET-II}

\author{Yiheng Ye\thanks{yihengy@slac.stanford.edu}, Spencer Gessner, SLAC National Accelerator Laboratory, Menlo Park, USA}
	
\maketitle

\begin{abstract}
    We present transverse phase space tomographic reconstruction of the two-bunch configuration at FACET-II. We demonstrate diagnosing two superimposed electron bunches with different transverse phase spaces using a single quadrupole scan. Tuning the bunch charge revealed unexpected space-charge-dominated dynamics.
\end{abstract}
\begin{figure*}
    \centering
    \includegraphics[width=\linewidth]{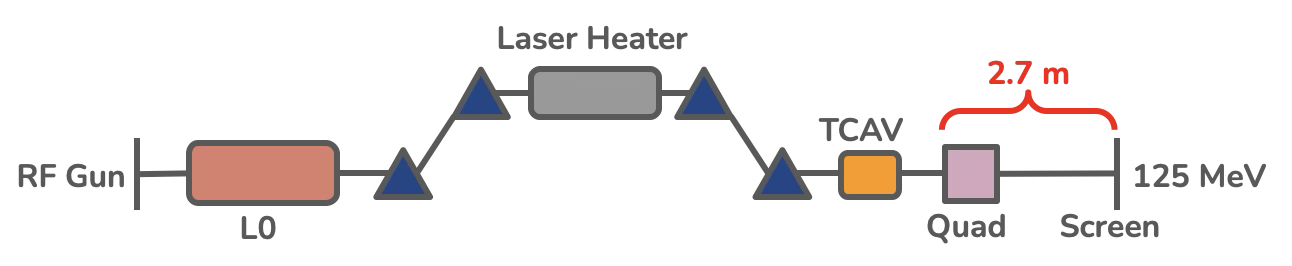}
    \caption{Schematics of the FACET-II injector. The quadrupole magnet and the screen we use for quadrupole scan is at the end of the injector sector with a distance of roughly 2.7 m.}
    \label{fig:injschem}
\end{figure*}

\section{Introduction}
Emittance characterizes beam quality in accelerators and is typically measured via quadrupole scans: we scan a quadrupole magnet upstream and record beam sizes on a downstream screen, then quadratically fit beam sizes versus quadrupole strength to extract transverse emittance $\epsilon$ and Twiss parameters $\alpha$, $\beta$, and $\gamma$ (see e.g. \cite{minty2003measurement}). This method assumes gaussian beam profiles, which often don't reflect actual operating conditions. With non-gaussian beams, this measurement technique may be inaccurate and provide minimal insights for accelerator tuning.

Transverse phase space reconstruction at FACET-II is based on tomography \cite{mckee1994phase}: reconstructing $n$-dimensional objects from $(n-1)$-dimensional projections at different angles. In quadrupole scans, the $x(y)$-profile at the screen represents a projection of $x(y)$-phase space at a specific angle for each quadrupole setting. We use detailed screen profiles from quadrupole scans to reconstruct the complete transverse beam phase space.

The two-bunch mode at FACET-II is used to generate drive and trailing bunches for plasma wakefield experiments~\cite{PhysRevAccelBeams.22.101301}. The injector laser is split into double pulses to generate driver-witness electron bunch pairs, which are then co-accelerated through the linac in the same rf bucket. In nominal operating mode, the driver-to-witness bunch charge ratio is 3:1 with a time separation of around 7 ps. Both the bunch charge and the time separation can be tuned to satisfy the needs of different experiments. The transverse phase spaces of the two bunches overlap, but are not identical. The combined two-bunch phase space is generally non-gaussian.

We have implemented tomographic measurements in the FACET-II injector in order to characterize the two-bunch phase space. Our implementation is based on references~\cite{hock2011, hock2013maximum, awake_tomography_gitlab}. We scanned the bunch charge ratio to observe transverse space-charge dynamics in the injector and tuned the bunch separation to observe effects of bunch generation in different rf phases. Our experimental observations diverge from simulation results, indicating that follow-up studies will be required.


\section{Transverse Phase Space Tomography}

Transverse tomography is carried out in FACET-II's injector beamline, using a quadruple and screen separated by 2.7 m as shown in Fig.~\ref{fig:injschem}. Given a quadruple strength $k$ and the drift length $d$, we can calculate the transfer matrix $M(k)$ between the quadruple and the screen. The coordinates of an electron in the $x(y)$ phase space are given by
\begin{equation}
    \begin{pmatrix}
x_{\text{screen}} \\
x^\prime_{\text{screen}}
\end{pmatrix} = \begin{pmatrix}
    M_{1,1}(k) & M_{1,2}(k)\\
    M_{2,1}(k) & M_{2,2}(k)
\end{pmatrix}\begin{pmatrix}
x_{\text{quad}} \\
x^\prime_{\text{quad}}
\end{pmatrix}.
\end{equation}
The quadruple magnet both rotates and shears the phase space of the beam, such that in each plane the projection of the image on the screen is the projection of the phase space at angle $\theta$, scaled by a factor of $s$, where $\theta$ and $s$ are given by
\begin{equation}\label{ang_scal_form}
    \theta = \tan^{-1}\left(\frac{M_{1,1}(k)}{M_{1,2}(k)}\right),\, s = \sqrt{M^2_{1,1}(k) + M^2_{1,2}(k)}.
\end{equation}

Projections of the beam image at the screen are stacked together to form a sinogram. Then it is rescaled and labeled according to equation(\ref{ang_scal_form}). We reconstruct the phase space using Maximum Likelihood Expectation Maximization (MLEM) algorithm \cite{shepp1982mlem}.

\section{Two bunch Configuration}

In the two-bunch configuration, the default quad scan that fits scalar beam sizes provides limited information on the transverse phase space. The drive bunch and the witness bunch are co-accelerated in the same rf bucket at different rf phases determined by their separation in time. They may have different Twiss parameters at a given location because of the energy difference between the two bunches.

\begin{figure*}[!htb]
    \centering
    \includegraphics*[width=1\columnwidth]{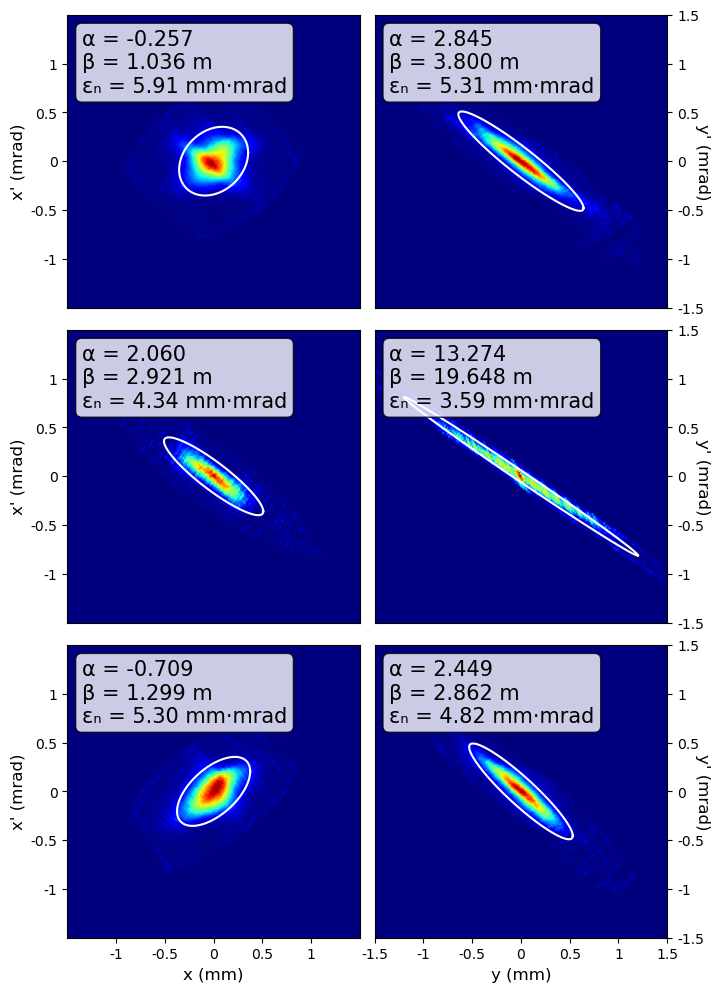}
    \hfill
    \includegraphics*[width=1\columnwidth]{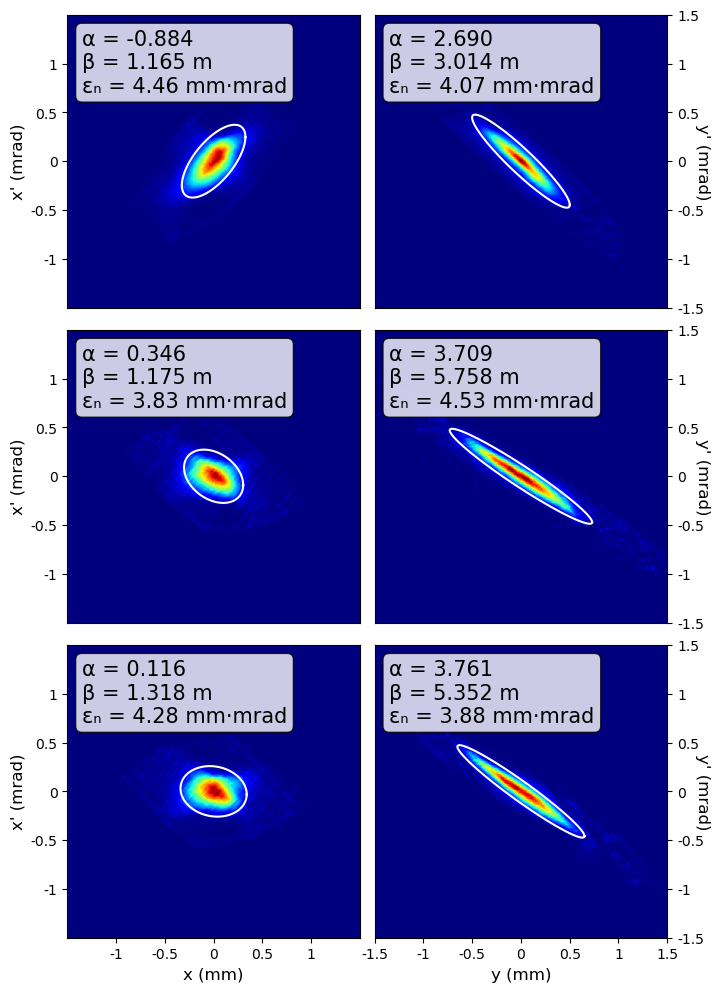}
    \caption{Tomographically reconstructed transverse phase space of two-bunch configurations in the FACET-II injector. Left: Nominal two bunch configuration showing x- and y-phase space of two bunches (first row), witness bunch (second row), and driving bunch (third row) with white ellipses from standard quadratic fitting ($3\varepsilon$). Right: Equal charge configuration where the injector laser is split into two beams with equal intensity and recombined with zero time separation.}
    \label{fig:2bunch}
 \end{figure*}

The nominal two bunch configuration for the plasma wakefield acceleration (PWFA) experiment consist of a 1.6 nC drive bunch and a 0.5 nC witness bunch. We did a quadruple scan in the injector to analyze the transverse phase space. Figure~\ref{fig:2bunch} (left) shows the tomographically reconstructed transverse phase spaces of the nominal two-bunch setting in the injector. We observe an X-shaped pattern in the horizontal phase space, which we interpret as the superposition of the two horizontal phases spaces of the individual bunches. This is confirmed by characterizing each bunch individually.


We set the laser pulse separation to 0 and observe the effect of bunch charges. Figure \ref{fig:2bunch} (right) shows the "two bunch", "witness", and "drive" phase spaces. In this case, the "two bunch" is a single bunch with 2.1 nC of charge, while "witness" and "drive" bunch are approximately the same with equal phase and charge. In both plane, phase spaces are stretched out in the spatial direction. We can see that the x phase space is flipped when the bunch charge is halved.

\begin{figure}[htbp]
    \centering
    \includegraphics[width=\linewidth]{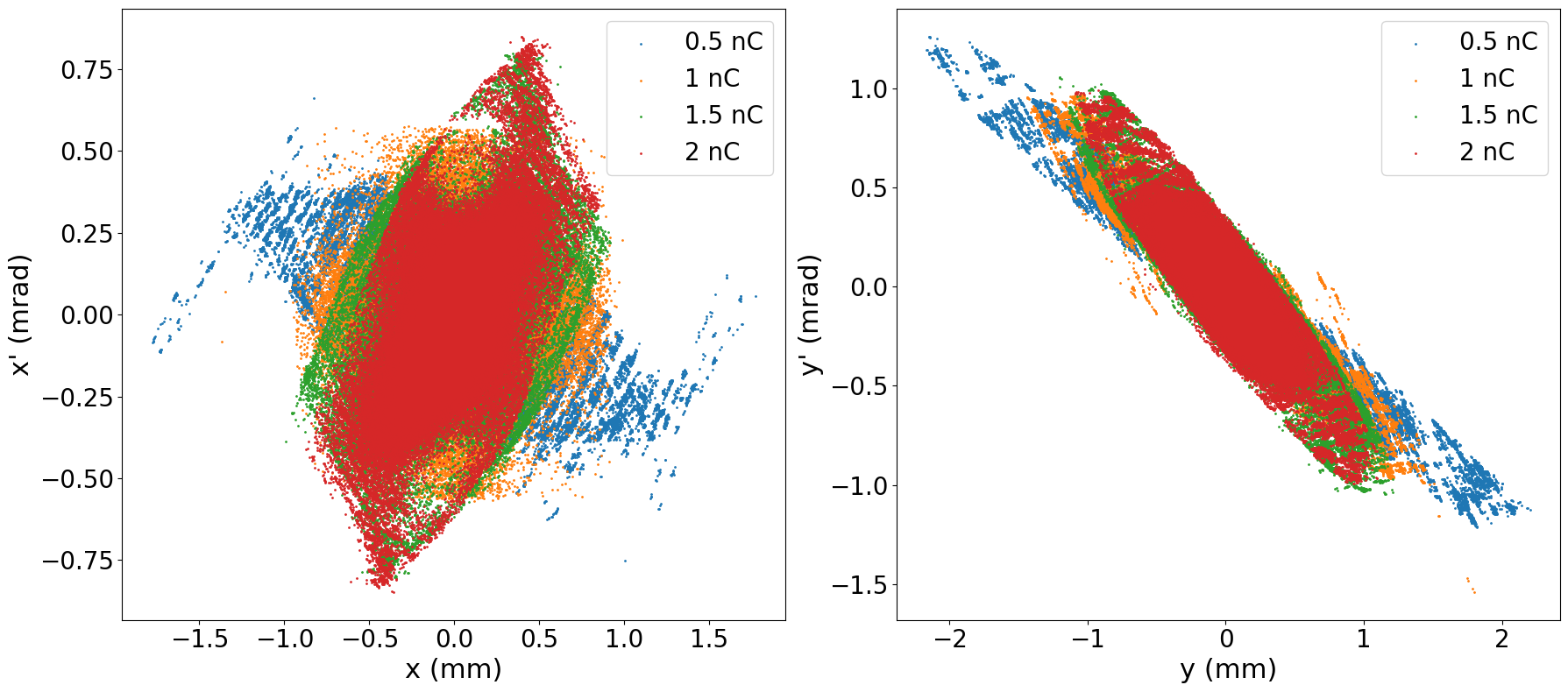}
    \caption{By splitting the injector laser into 1:1 and 3:1, respectively, and measure the phase space at the same location, we map out transverse phase spaces with different bunch charges. The scatter plot is obtained by sampling from the reconstructed phase space distribution in each case.}
    \label{fig:combined-phase-space}
\end{figure}

We further mapped out the phase space with roughly 0.5 nC, 1 nC, 1.5 nC, and 2 nC bunch charge, respectively, and the result is shown in figure \ref{fig:combined-phase-space}. We observe that the higher the bunch charge, the more spatially stretched the phase space is.

In the horizontal plane, the result does not agree with our preliminary simulation result. We attributed the mismatch to three effects and their interplay: (1) space-charge force; (2) solenoid setting for emittance compensation; and (3) collective effects in the laser heater chicane. During the experiment, the solenoid setting is unchanged, so it is not efficiently compensating bunches with different charges. Collective effects in the laser heater chicane can play a major role, since the horizontal phase space is particularly affected. A comprehensive simulation and analysis are in progress to understand the measurement result.

\begin{figure}[htbp]
    \centering
    
    \includegraphics[width=0.5\textwidth]{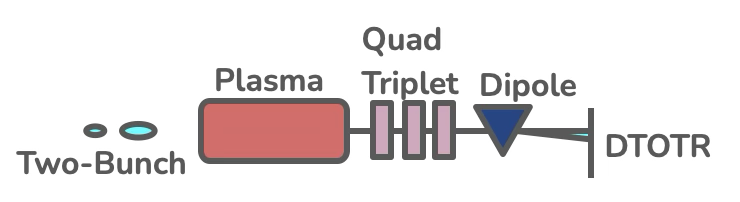}
    \caption*{(a)}
    
    \vspace{0.3cm} 
    
    \includegraphics[width=0.155\textwidth]{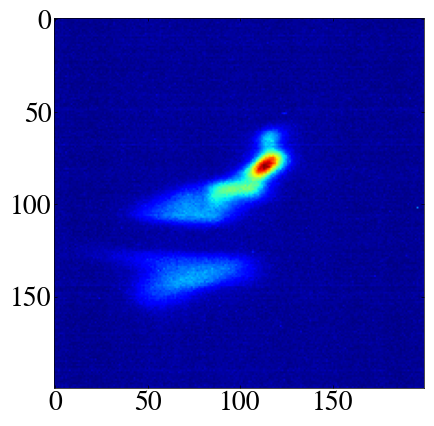}
    \includegraphics[width=0.155\textwidth]{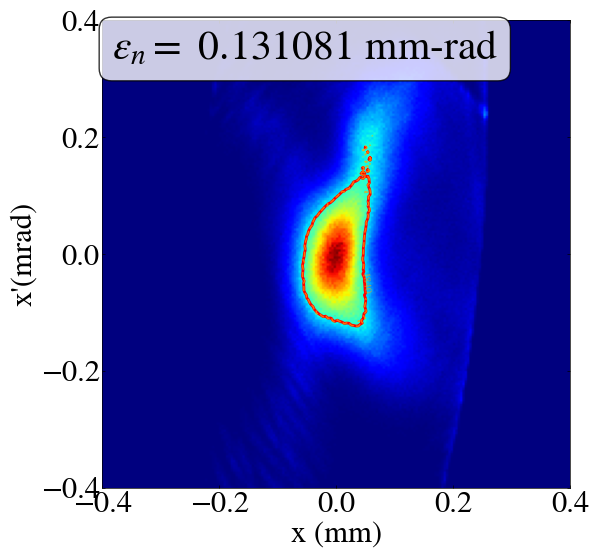}
    \includegraphics[width=0.155\textwidth]{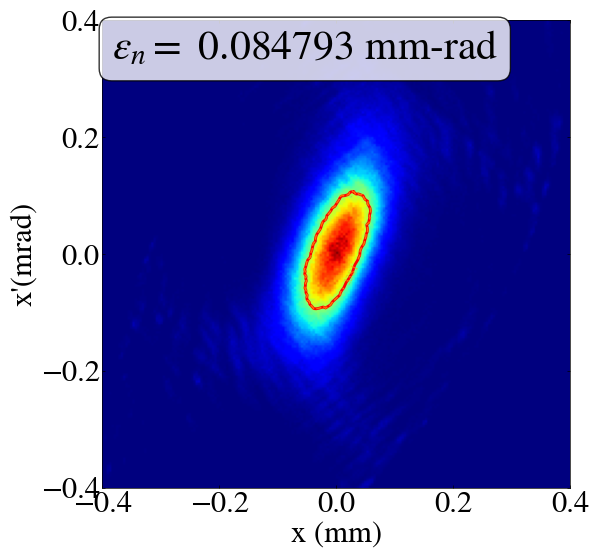}
    \caption*{(b)}
    
    \caption{(a) Schematics of the FACET-II spectrometer beamline. (b) Preliminary incoming beam phase space reconstruction results. Left: Example of a DTOTR screen image. Middle and right: reconstructed phase space in the non-dispersive plane for drive and witness bunches, respectively. The top 39\% of pixels were selected for emittance calculation (equivalent to Gaussian beam definition). During two-bunch cathode operation, the linac was optimized for the witness bunch at the expense of the drive bunch's quality.}
    \label{fig:spec_tomo}
\end{figure}

\section{Future Directions}

In this section, we briefly mention some preliminary results and future applications of tomographic phase space reconstruction method.

Among other things, the FACET-II spectrometer beamline \cite{PhysRevAccelBeams.27.051302} consists of three quadruple magnets, a dipole magnet for dispersion, and an optical transition radiation (OTR) screen
[Figure \ref{fig:spec_tomo}(a)shows the schematic of the spectrometer beamline]. It is the final section of FACET-II that is used for energy-resolved beam diagnostic. These quadruple magnets can be used to perform quadruple scans, from which we can reconstruct the phase spaces in the non-dispersive plane at any location upstream. Some preliminary results are shown in figure \ref{fig:spec_tomo}(b). These are phase spaces of the beam without plasma (i.e. in vacuum). We plan to utilize this technique to better characterize the incoming beam for PWFA experiments. In addition, we also plan to implement single-shot reconstruction of the plasma-accelerated beam phase space from the so-called "butterfly plot" \cite{cooke2024emittance}. 

The most important application will be the use of this tool to study the beam dynamics in the experimental area of FACET-II. The ultimate goal is to tune the drive beam and the witness beam so that the acceleration occurred when both bunches have high beam quality and a matched beta function.

\section{ACKNOWLEDGEMENTS}
We thank N. Majernik and D. Storey for insightful discussions. This work was supported by the U.S. Department of Energy under DOE Contract No. DE-AC02-76SF00515.

%
%
\ifboolexpr{bool{jacowbiblatex}}%
	{\printbibliography}%
	{%
	

} 
%
%


\end{document}